\documentstyle[aps,prl,psfig,floats]{revtex}
\begin{document}
\title{Quantum Decoherence and Weak Localization at Low Temperatures}

\author{Dmitrii S. Golubev$^{1,3}$ and Andrei D. Zaikin$^{2,3}$}
\address{$^{1}$ Physics Department, Chalmers University of Technology,
S-41296 G\"oteborg, Sweden\\
$^2$ Institut f\"{u}r Theoretische Festk\"orperphysik,
Universit\"at Karlsruhe, 76128 Karlsruhe, Germany\\
$^3$ I.E.Tamm Department of Theoretical Physics, P.N.Lebedev
Physics Institute, Leninskii pr. 53, 117924 Moscow, Russia}

\maketitle

\begin{abstract}
We discuss a fundamental effect of the interaction-induced
decoherence of the electron wave function in disordered metals.
In the first part of the paper we consider a simple model of a
quantum particle interacting with a bath of harmonic oscillators
and analyze the physical origin of the effect. This exactly solvable
model also allows to understand why the arguments against the
existence of the effect at low temperatures fail. The second
part of the paper is devoted to a rigorous analysis of
quantum decoherence in disordered metals. We also
discuss the relation of our results to the recent experiments
on GaAs structures. The existence of a finite quantum decoherence rate
at low $T$ implies that low dimensional disordered metals with
generic parameters do not become insulators even at $T=0$.
\end{abstract}

\section{Introduction}

The concept of quantum coherence is one of the
most fundamental in quantum mechanics. According to the general
principles quantum coherence of the wave function cannot be
destroyed due to {\it elastic} interaction with a {\it static}
external potential. On the other hand, {\it inelastic} scattering
processes may (and in general do) destroy the phase coherence
of the wave function. For electrons in a metal
such processes (e.g. inelastic electron-electron and electron-phonon
scattering) are important at sufficiently high temperatures. But
as the temperature $T$ is lowered inelastic processes become
less intensive
and the corresponding inelastic scattering time $\tau_i$ for
an electron in equilibrium tends to infinity at $T \to 0$.
Hence, one could consider natural that interaction may cause
dephasing in a metal at nonzero $T$, but {\it not at} $T=0$.

Surprizingly, it was found in many experiments with
disordered metals (see \cite{Webb} and references therein)
that the
effective dephasing time $\tau_{\varphi}$ for the electron wave
function saturates at the level $\tau_{\varphi} \sim 10^{-1}\div 10$ ns and does
not depend on temperature below $T \sim 1$K. These findings are
in a {\it clear contradiction} with the above point of view
according to which the time $\tau_{\varphi}$ should increase
to infinity as $T$ approaches zero. Scattering on magnetic
impurities, heating and the external noise were suggested
as possible reasons for saturation of $\tau_{\varphi}$
observed in earlier experiments. All these reasons were
convincingly ruled out by the authors \cite{Webb}, and it was
argued that it is the {\it zero point motion of electrons} in
a disordered conductor that causes dephasing at $T=0$.

A theory of the effect of quantum decoherence at low
temperatures was suggested by the present authors \cite{GZ}.
It was demonstrated that at sufficiently low $T$
{\it the high frequency quantum noise} of the effective electronic
environment is responsible for the effect of quantum decoherence.
It was predicted that at low $T$ the decoherence rate
$1/\tau_{\varphi}$ for 1d metallic systems saturates at the level
\begin{equation}
1/\tau_{\varphi}\simeq e^2v_F/\pi \sigma_1,
\label{1}
\end{equation}
where $\sigma_1$ is the wire conductance per unit length, $v_F$
is the Fermi velocity. The result (\ref{1}) is in a very good
agreement with the experimental findings \cite{Webb} as well as
with previous experimental results.

The predictions \cite{GZ} inevitably lead to another
{\it fundamental conclusion}: low dimensional metals
{\it do not become insulators even at} $T=0$. Indeed, e.g.
in the case of 1d systems the localization length is
$l_{\rm loc} \sim N_{\rm ch}l$, while the result (\ref{1}) yields
the effective decoherence length $L_{\varphi} \sim l\sqrt{N_{\rm ch}}$,
where $l$ is the elastic mean free path and $N_{\rm ch}$ is the
effective number of conducting channels.
It is obvious that for $N_{\rm ch} \gg 1$ (which is always the case
in metallic wires) $L_{\varphi}$ is {\it parametrically}
smaller than $l_{\rm loc}$. This implies that due to
interaction with other electrons the phase coherence of the
electron wave function is destroyed {\it faster} than the
electron can get localized, i.e. localization {\it never} takes
place under the above conditions.

Since the results \cite{Webb,GZ} imply a necessity to strongly
reconsider the commonly adopted point of view on the role
of interaction in disordered metals at low $T$, it is not
surprizing that the subject still remains controversial.
Moreover, it is sometimes argued that quantum decoherence
at $T=0$ would contradict to general principles of quantum
mechanics. What are the arguments against the quantum noise in a
disordered metal as a reason for quantum decoherence of electrons?
Usually it is argued on a general level,
adopting a much simpler quantum mechanical model in order
to illustrate the main statement. On one hand this makes sense
indeed: the nature of a fundamental quantum
mechanical effect (or the absence of it) can and should be
understood without unnecessary complications. On the other
hand, one has to make sure that all significant features which
yield the effect in a more complicated problem are still
present in a simplified one. There is an obvious danger in
making too many simplifications: one can easily throw out
the baby with the bathwater.

Consider a quantum particle colliding with a harmonic
oscillator with the frequency $\omega$.
Assume that the interaction potential is short range and before the collision
the particle was in the state with the energy {\it smaller} than
$\hbar \omega$, while the oscillator was in the ground state.
The general features of this
{\it scattering problem} are well known: after interaction
the oscillator remains in the ground state (the particle does not have
enough energy to excite it), the particle energy also
remains the same as before interaction, the particle wave function
-- although in general changes as a result of interaction --
stays fully coherent. Thus no energy exchange between the particle
and the oscillator has occured, and no phase coherence has been lost
due to interaction.
What remains is to understand if
the above consideration is relevant to the problem in question.

At the first sight it appears
to be relevant indeed. Although in a metal the electron interacts with
other electrons instead of oscillators (let us ignore the
electron-phonon interaction for simplicity), at $T=0$ all states
below/above the Fermi energy should be occupied/empty, and no
energy can be transferred: the electron cannot be scattered into
any lower energy state due to the Pauli principle
as well as to any higher energy state because no energy can be
extracted from other electrons due to the same reason.
It follows immediately, that at $T \to 0$ the phase coherence of the
electron wave function in a metal cannot be lost due to the very same
reason as in the above example with the particle and the oscillator:
there exists no inelastic scattering.

This consideration turns out to be too naive, however. It may be
sufficient if the {\it infinite} time limit can be taken already
at the very first step of the calculation as e.g. within the
standard formulation of the scattering problem. In this case the
result will be proportional to the delta-function of the energy
difference between the initial and the final states ensuring
the energy conservation. However, at any finite time the
energy-time uncertainty principle should be taken into account.
The energy uncertainty can be not necessarily small, especially
if the particle interacts with an {\it infinite} number of oscillators.
This is crucially important for the effect in question.
On top of that our problem is more complicated because we are dealing
with an {\it interacting} quantum many body system rather than the
scattering problem for two particles with well defined in- and out-states.
The electron energy fluctuates as a result of interaction with other
electrons, only the total energy of all electrons plus the interaction
energy is conserved. One may introduce the effective oscillators also in this
case, they are just the modes of the electromagnetic field. The interaction
between the electron and one such mode can be naively described as
$H_{\rm int}=eE\cos({\bf kx})$, where $E$ is the amplitude of the electric
field, ${\bf k}$ is the wave vector and ${\bf x}$ is the electron coordinate.
The interaction potential remains finite for any ${\bf x}$, i.e.
the electron always interacts with the oscillator and
the scattering problem cannot be formulated. Actually the
problem is even more complicated because the electron interacts
with an infinite number of such oscillators at the same time.
It is also important that {\it in the presence of interaction} the energy
exchange between the low energy particle and the high energy oscillator is
possible {\it without excitation} of the latter simply because its lowest
energy level acquires a finite width. This effect depends on
the strength of interaction, but it always exists because the interaction is
never ``turned off''.

One might argue that this simply means that electrons are ``bad''
particles in the presence of interaction and one should rather
define ``better behaving'' quasiparticles and calculate all
measurable quantities in their terms. This does not always help
because of at least two reasons. One is that quasiparticles (if they exist)
are usually defined within an approximate procedure. Although
the approximation may work well for calculation
of certain physical quantities, it may fail for other quantities.
Another reason is that the transformation of the basis (even if it can
be done exactly) may not be convenient if the measurements are done
only with a ``bad'' particle. In this case calculation with ``good''
quasiparticles can be by far more complicated, and it is much simpler
to ``get rid'' of all but one ``interesting'' degree of freedom at an early
stage of calculation by tracing them out in the full density matrix.
This is the key idea of the Feynman-Vernon theory
of the influence functionals \cite{FV,FH} developed further
by Caldeira and Leggett \cite{CL} and Schmid \cite{Schmid1} in
application to an infinite bath of harmonic oscillators with the ohmic
spectrum.

Superconducting Josephson junctions may serve as an example of a
fermionic system where the same ideas have been worked out \cite{AES,SZ}.
In this case it is possible to exactly integrate out all electron
degrees of freedom and describe the system dynamics in terms of only
one collective variable -- the Josephson phase $\varphi$. Of course,
$\varphi$ is known to behave ``badly'' (its quantum dynamics is
incoherent in almost all cases, see e.g. \cite{SZ,Schmid2,ChLeg})
but namely this variable is of interest in
Josephson junctions and SQUIDs just because the junction current and
voltage operators as well as the flux operator in SQUIDs are defined in
terms of $\varphi$. In this case ``better behaving'' quasiparticles
are simply irrelevant. Therefore it appears to be unreasonable
to argue in favour of ``coherent'' quasiparticles if all measurements
are done only with ``incoherent'' variables.
We believe that a similar situation is encountered
if one calculates the conductance of a disordered metal. Actually
in this case it is even more complicated because it is not clear if
``good'' quasiparticles can be introduced at all.

The paper is organized as follows. In Section 2 we will
demonstrate how the fundamental effect of quantum decoherence can
be derived within a simple model of a quantum mechanical particle
interacting with an environment consisting of a collection of
harmonic oscillators. It is remarkable that already this simple
model captures all significant features of the effect and allows
to understand why the arguments against its existence fail.
We also establish the relation between our results and the
standard perturbative treatment of a scattering problem.
Section 3 is devoted to a rigorous
microscopic analysis of quantum decoherence in a disordered metal.
Discussion of recent experiments and comparison with our theory are
presented in Section 4. The main conclusions are outlined in Section 5.

\section{Feynman-Vernon Theory and Quantum Decoherence}

\subsection{Influence Functionals}

Consider a quantum-mechanical particle (or a more general system)
characterized by a coodinate $q(t)$ and the
action $S_0[q(t)]$. Assume that this particle interacts with another
quantum system described by a coordinate $Q(t)$ and the action $S_{\rm env}(Q)$.
We will call the latter quantum system ``environment''.
The total action for the system ``particle+environment'' has the form
\begin{equation}
S[q,Q]=S_0[q]+S_{\rm env}[Q(t)]+S_{\rm int}[q(t),Q(t)],
\label{SqQ}
\end{equation}
where $S_{\rm int}$ describes interaction between $q$ and $Q$.

Suppose we are interested in the probability $W$ for the particle $q$
to have the coordinate $q=q_f$ at a time $t$ provided
at $t=0$ it had the coordinate $q=q_i$. A general and elegant way to
solve this problem can be formulated within the Feynman-Vernon theory of
the so-called influence functionals. An extensive discussion of
this theory can be found in Refs. \cite{FV,FH}. Here we only repeat
the key steps.

By definition the probability $W$ is given by the square of a transition
amplitude $W=|K(q_f,t;q_i,0)|^2$. In the absence of interaction ($C=0$)
the oscillator coordinate $Q$ does not enter into the expression for $K$.
However for $C \neq 0$ there appears an additional force acting on a
particle $q$. This force depends on $Q$ which is itself a quantum variable.
Thus we cannot anymore restrict ourselves to the dynamics of $q$, but
rather should deal with the total Hamiltonian for a system ``$q+Q$'' or,
equivalently, with a complete action (\ref{SqQ}). Of course, the
probability $W$ is again equal to the square of the corresponding transition
amplitude $K$ which now depends on both $q$ and $Q$. It is important,
however, that we are not interested in the final state of the
oscillator and do not make any measurements of $Q$. Therefore we
should sum over {\it all} possible final states of $Q$ and the formula
for the probability $W$ takes the form
\begin{equation}
W=\sum_{Q_f}K(q_f,Q_f,t;q_i,Q_i,0)K^*(q_f,Q_f,t;q_i,Q_i,0).
\label{W1}
\end{equation}
This is an important point. Each of the terms in the sum (\ref{W1})
represents a probability for a system to come into the state $(q_f,Q_f)$.
Since the subsystem $Q$ can be in {\it any} final state $Q_f$, in order
to find the total probability $W$ we have to add all these probabilities
together.

One can slightly generalize the problem and describe the evolution of
the density matrix of the system $\rho (q,q')$ from some initial to some
final state. This evolution is described by the equation
\begin{equation}
\rho (q_f,q'_f)=
\int {\rm d}q_i{\rm d}q'_iJ(q_f,q'_f,t;q_i,q'_i,0)\rho_i(q_i,q'_i),
\label{rho1}
\end{equation}
where $\rho_i(q_i,q'_i)$ is the initial density matrix of the particle $q$.
For the sake of simplicity in what follows we will assume that there is
no interaction between $q$ and $Q$ before $t=0$. Then the total initial
density matrix can be factorized as $\rho_i(q_i,q'_i)\tilde\rho_i (Q_i,Q'_i)$.
Here the kernel $J$ is again given by the product of two amplitudes. It is
convenient to represent this product in terms of a double path integral
\begin{equation}
J=\int_{q_i}^{q_f}{\rm D}q_1\int_{q'_i}^{q'_f}{\rm D}q_2
\exp ({\rm i}S_0[q_1(t)]-{\rm i}S_0[q_2(t)])F[q_1(t),q_2(t)],
\label{J}
\end{equation}
where $F$ is the {\it influence functional} which describes the
total effect of the subsystem $Q$ on the particle $q$. The functional
$F$ in turn can be represented in terms of a double path integral
\begin{eqnarray}
F[q_1(t),q_2(t)]&=&\sum_f\int {\rm d}Q_i\int {\rm d}Q'_i\int_{Q_i}^{Q_f}
{\rm D}Q_1
\int_{Q'_i}^{Q'_f}{\rm D}Q_2\tilde\rho (Q_i,Q'_i)\times
\nonumber \\
&&
\exp ({\rm i}S_{\rm env}[Q_1]-{\rm i}S_{\rm env}[Q_2]+
{\rm i}S_{\rm int}[q_1,Q_1]-{\rm i}S_{\rm int}[q_2,Q_2]).
\label{F}
\end{eqnarray}
Here again (we cite from \cite{FH}) ``$\sum_f$ just means that at
some final time $t_f$ after we are no longer interested in the
interaction we must take $Q_f=Q'_f$ and integrate over all $Q_f$'',
i.e.
\begin{equation}
\sum_f \to \int {\rm d}Q_f\int {\rm d}Q'_f\delta (Q_f-Q'_f).
\label{sum}
\end{equation}
This completes the general analysis of Feynman and Vernon.
Its important advantage is that no approximations
have been done so far, i.e. the above formulas are {\it exact}.
One more citation from \cite{FH} is in order:
``$F$ contains the entire effect of the environment including
the change in behavior of the environment resulting from reaction
with $q$. In the classical analogue, $F$ would correspond to knowing
not only what the force is as a function of time, but also what it
would be for every possible motion $q(t)$ of the object. The force
for a given environmental system depends in general on the motion
of $q$, of course, since the environmental system is affected by
interaction with the system of interest $q$''. Thus all changes
of the state of the environment resulting from the interaction
with a dynamical variable $q$ are {\it automatically} taken
into account within the above formalism.

In order to proceed we first assume that the environment consists
of a single harmonic oscillator $Q$ which has a unit mass and a
frequency $\omega$. For simplicity, the interaction is chosen bilinear with
respect to both the particle and the oscillator coordinates $q$ and $Q$,
so that the total action for the system ``particle+oscillator'' has the form
\begin{equation}
S[q,Q]=S_0[q]+\int {\rm d}t\left( \frac12\dot Q^2 -\frac{\omega^2}{2}Q^2+
CqQ\right),
\label{SqQ1}
\end{equation}
where $C$ is a constant which governs the strength of interaction. We will
also assume that the oscillator is initially kept at a temperature $T$, i.e.
the probability to occupy the state $k$ is
$w_k= \exp (-k\omega /T)(1-\exp (-\omega /T))$.
In the limit $T \ll \omega$ the initial state of the oscillator is
its ground state $k=0$.

For this model it is a matter of a simple integration over the
$Q$-variables to obtain the exact expression for the influence
functional $F$ (\ref{F}). One finds \cite{FV,FH}:
\begin{equation}
F[q_1,q_2]=\exp (-{\rm i}S_R[q_1,q_2]-S_I[q_1,q_2]).
\label{F2}
\end{equation}
Defining $q_+=(q_1+q_2)/2$ and $q_-=q_1-q_2$ we have
\begin{equation}
S_R=\frac{C^2}{\omega}\int_{0}^{t}{\rm d}t_1{\rm d}t_2q_-(t_1)
\sin (\omega (t_1-t_2))q_+(t_2),
\label{SR0}
\end{equation}
\begin{equation}
S_I=\frac{C^2}{2\omega}\coth \left( \frac{\omega}{2T}\right)
\int_{0}^{t}{\rm d}t_1{\rm d}t_2q_-(t_1)
\cos (\omega (t_1-t_2))q_-(t_2).
\label{SI0}
\end{equation}
Let us point out that $S_I\geq 0$ for {\it all} trajectories $q(t)$.
Eqs. (\ref{F}-\ref{SI0}) summarize the complete effect of interaction
with the oscillator $Q$ on quantum dynamics of a particle $q$.

\subsection{Free particle interacting with oscillators}

Let us come back to the probability $W$ (\ref{W1}) or,
more generally, to the kernel $J$ (\ref{J}). We will consider two simple
examples.
The first example is a freely propagating quantum particle with a mass $m$.
In the absence of interaction ($C=0$) we have $F \equiv 1$, and the
double path integral (\ref{J}) decouples into the product of two
single integrals. For $q_{i/f}=q'_{i/f}$ each of them is dominated by
{\it the same} classical path $\tilde q(t')=q_i+(q_f-q_i)t'/t$
and (cf. e.g. \cite{FH})
\begin{equation}
K(q_f,t;q_i,0)= \sqrt{m/2\pi {\rm i}t} \exp ({\rm i}S_0[\tilde q]),
\;\;\; S_0[\tilde q]=m(q_f-q_i)^2)/t.
\label{amp}
\end{equation}
The probability $W=KK^*$
does not depend on the phase ${\rm i}S_0$ of each of the amplitudes,
these phases enter with the opposite signs and cancel.

Now let us turn on the interaction ($C\neq 0$). If -- just for
the sake of simplicity -- one treats this interaction perturbatively, one
immediately observes that the double path integral (\ref{J}) is again
dominated by the same classical path $q_1(t')=q_2(t')=\tilde q(t')$.
For {\it any} $q_1(t')=q_2(t')$ we have $S_R=S_I\equiv 0$ and therefore
$F[q,q]\equiv 1$ like in the absence of interaction.
One can also establish the general form of the kernel $J$ (\ref{J}).
It is relatively complex and is not presented here. More interesting
situation emerges if we modify our model and
consider our particle $q$ interacting with $N$ oscillators
with frequencies $\omega_n$. The corresponding generalization
is trivial: one should just substitute $\omega \to \omega_n$
in (\ref{SRI1}) and carry out the summation over all $1 \leq n \leq N$.
If one sends $N$ to infinity and assumes a continuous
distribution of the oscillator frequencies \cite{CL,Schmid1}:
\begin{equation}
\sum_{n}\frac{\pi C^2}{2\omega_n}[\delta (\omega_n-\omega )
-\delta (\omega_n+\omega )]= \eta \omega ,\;\;\;\; |\omega | <\omega_c,
\label{ohm}
\end{equation}
($\omega_c$ defines the high frequency cutoff) one arrives at the
influence functional of the form (\ref{F2}-\ref{SI0}) where
one should substitute
\begin{equation}
\frac{C^2}{\omega}(...) \to \eta\int_{-\omega_c}^{\omega_c}
\frac{\omega d\omega}{2\pi} (...).
\label{subs}
\end{equation}

The problem defined by eqs. (\ref{F2}-\ref{SI0},\ref{subs}) is
gaussian and can be solved exactly, see e.g. \cite{CL2}.
Performing a straightforward gaussian integration over $q$
one arrives at the the exact expression for the kernel $J$ (\ref{J}):
\begin{eqnarray}
J&=&\frac{\eta}{2\pi(1-{\rm e}^{-\gamma t})}
\exp\bigg[{\rm i}\eta\frac{q_{+f}q_{-f}+q_{+i}q_{-i}
-{\rm e}^{\gamma t}q_{+f}q_{-i}-q_{-f}q_{+i}}{{\rm e}^{\gamma t}-1}
\nonumber\\
&&
-mf_1(t)q_{-i}^2-mf_2(t)(q_{-f}-q_{-i})^2-
mf_3(t)q_{-i}(q_{-f}-q_{-i})
\bigg].
\label{Jint}
\end{eqnarray}
Here $\gamma=\eta/m$, $q_{\pm i/f}$ are initial/final values of $q_{\pm}$,
\begin{eqnarray}
f_1(t)&=&\frac{\gamma}{2}\int\limits_0^t{\rm d}s \int\limits_0^t{\rm d}s'
\int\limits_{-\omega_c}^{\omega_c}\frac{{\rm d}\omega^2}{4\pi}
\coth\frac{\omega}{2T}{\rm e}^{-{\rm i}\omega (s-s')}=
\gamma Tt+\gamma \ln\frac{1-{\rm e}^{-2\pi Tt}}{2\pi (T/\omega_c)}
\label{f1}
\end{eqnarray}
and the functions $f_2(t)$ and
$f_3(t)$ tend to the following values in the interesting limit of long times:
\begin{equation}
f_2=\left\langle\frac{m\dot q^2}{2}\right\rangle=
\gamma\int\limits_0^{\omega_c}\frac{{\rm d}\omega}{2\pi}
\frac{\omega\coth\frac{\omega}{2T}}{\omega^2+\gamma^2}
\simeq
\frac{\gamma}{2\pi}\ln\frac{\omega_c}{\gamma}+
\frac{T}{\pi}\arctan\frac{T}{\gamma}.
\label{energy}
\end{equation}
and $f_3=T+2f_2$. The obtained exact solution allows to make several
important observations. One of them is completely obvious: the particle $q$
looses its coherence due to interaction with the Caldeira-Leggett
bath of oscillators. Indeed in the long time limit we have $f_1(t)\gg f_2$
and the kernel (\ref{Jint}) effectively reduces to
\begin{equation}
J=\frac{1}{2}\sqrt{\frac{\eta \gamma}{\pi f_1(t)}}{\rm e}^{-mf_2q_{-f}^2}
\delta(q_{-i})
\to
\frac{1}{L}{\rm e}^{-m f_2q_{-f}^2}\delta(q_{-i}),
\label{limit}
\end{equation}
where $L$ is the system size. In other words, for any initial conditions the
density matrix tends to the same equilibrium form
$\rho(q_1,q_2)=(1/L){\rm e}^{-m f_2(q_1-q_2)^2}$ which is not sensitive
to the initial phase. According to eqs.
(\ref{Jint},\ref{f1}) the decay of off-diagonal elements of the
initial density matrix is exponential at any nonzero $T$ with the
characteristic time $\tau_0 = 1/\eta T q_{-i}^2$. At $T=0$
the off-diagonal elements decay as a power law
\begin{equation}
\rho_i(q_{_i}) \propto (t\omega_c)^{-\eta q_{-i}^2},
\label{power}
\end{equation}
but also in this case the information about the initial phase
is practically lost in the long time limit.

Another observation is that at sufficiently long times
the average value of the kinetic energy
of the particle $m\dot q^2/2$ (\ref{energy}) is not zero even at $T=0$
{\it irrespectively} to its initial energy. At high temperatures
$T \gg \gamma \ln (\omega_c/\gamma )$ the energy is given by its classical
value $T/2$, but at lower $T$ its value is determined by the interaction
parameter $\gamma$ and the high frequency cutoff parameter $\omega_c$.
It is sometimes believed that if initially all the bath oscillators
are in the ground states and the particle energy is zero, no energy exchange
between the particle and the oscillators will be possible
because the particle has no energy to excite the oscillators and
the latter in turn cannot transfer their
zero-point energy to the particle. This statement is obviously incorrect
in the presence of interaction: the low energy particle will increase
its average energy while the interaction energy will be lowered to preserve
the energy conservation for the whole system.
The presence of $\omega_c$ in eq. (\ref{energy}) implies that there
exists the energy exchange between the particle and {\it all} oscillators
including the high frequency ones with $\omega \sim \omega_c$.
One should not think, however, that such oscillators need to be
excited in order to make this exchange possible. The
energy transfer mechanism is different: in the presence of interaction
the oscillator energy levels (including the ground state one) acquire
a finite width and the oscillator can exchange energy in arbitrarily
small portions.

It is important to emphasize that some of the above effects cannot be
correctly described within a naive perturbation theory in the
interaction based e.g. on the Fermi golden rule. Just for an
illustration let us choose the plane wave
$\psi \sim \exp (ip_1q)$ as the initial state of the particle $q$ and
evaluate the transition probability $W_{p_1p_2}$
to the state with the momentum $p_2$. Without interaction one has
$W_{p_1p_2}=\delta_{p_1p_2}$. For
small $\eta >0$ at $T=0$ in the long time limit one gets
from eqs. (\ref{rho1}), (\ref{Jint})
\begin{equation}
W_{p_1p_2} \sim \frac1{\sqrt{\chi_1\chi_2}}\exp \left(-\frac1{\eta}
\left( \frac{p_1^2}{\chi_1}+\frac{p_2^2}{\chi_2}\right)\right),
\label{Wpp}
\end{equation}
where $\chi_1 \simeq 4\ln (\omega_ct)$ and $\chi_2 \simeq
(2/\pi)\ln (\omega_c/\gamma )$. It is obvious that the result (\ref{Wpp})
cannot be recovered in any finite order of the perturbation theory in $\eta$.

The above model can also serve as an illustration of the role of
``good'' quasiparticles in the effect of quantum decoherence. It is
clear that in this model one can carry out exact diagonalization
of the Hamiltonian and introduce a new set of independent (and
therefore coherent) particles/oscillators. It is also clear that
this transformation will by no means influence our result for
the density matrix $\rho (q,q')$: this result is exact. Thus also
the calculation with ``good'' quasiparticles will yield
the incoherent dynamics of $q$, however with much more efforts
and with loss of physical transparency. The
basic reason for dephasing of $q$ is, however, transparent
also in this case: $q$ will be expressed as a sum of
{\it infinite} number of independent
particle/oscillator cordinates and therefore will be able to return to
its initial state only after infinite time.

\subsection{Particle on a ring}

The second example is a quantum particle on a ring. Again we would
like to calculate the probability $W$, but now we have to choose both
$q_i$ and $q_f$ on a ring. We choose $q_i=q_f=0$ (see Fig.1), i.e. $W$
is the probability for a particle to return to the same point $q=0$.
Again without interaction the two amplitudes $K$ and $K^*$ decouple
and can be evaluated separately. We have $K= \sum_mK_m$, where
$K_m \sim \exp ({\rm i}S_m)$ is the contribution from a path which traverses
$m$ times along the ring in a clockwise ($m>0$) or a counterclockwise
($m<0$) way and returns to the starting point $q=0$. It is obvious that
$S_m=S_{-m}=2\pi^2R^2m^2/t$, where $R$ is the ring radius.
If we neglect terms $K_mK^*_{m'}$ with $m' \neq \pm m$
(one can argue that e.g. for $S_1 \gg 1$ those are fast oscillating with
$m-m'$ terms which effectively cancel out in the course of summation
over $m$ and $m'$), then the probability $W$ can be written as a sum
of two terms $W=W_1+W_2$, where
\begin{equation}
W_1=\sum_mK_mK^*_m, \;\;\;\;\; W_2=\sum_mK_mK^*_{-m}.
\label{W12}
\end{equation}
The term $W_1$ is determined by a pair of equivalent paths $q_1(t')=q_2(t')$
(Fig. 1a). This contribution does not vanish in the classical
limit. The term $W_2$ comes from a pair of {\it time reversed} paths
$q_1(t')=q_2(t-t')$ (Fig. 1b). This term describes the effect of
quantum interference and therefore is very sensitive to the presence
of the phase coherence in our system. Obviously, $W_2$ vanishes in the
classical limit. As before, in both cases the
phase factors $\exp (\pm {\rm i}S_m)$ enter with opposite signs and cancel in
each of the terms in (\ref{W12}). Without interaction we have $W_2=W_1-|K_0|^2$.

\begin{figure}[h]
\centerline{\psfig{file=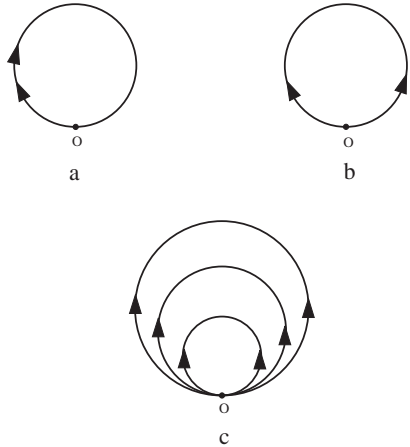,height=6cm}}
\caption{
A particle on a ring: (a) classical return paths, (b) time reversed return
paths, (c) time reversed return paths of different sizes.
}
\end{figure}

Now let us analyze the effect of interaction. The expression for
the return probability $W_1$ turns out to be insensitive to interaction
due to exactly the same reason as in our first example: this probability
is determined by the pairs of {\it equivalent} paths $q_1=q_2$ with
$F\equiv 1$. In contrast to $W_1$, the return probability $W_2$ determined
by the pairs of time reversed paths {\it is} affected by interaction.
We will restrict ourselves to the most interesting physical situation
when the energy of a particle $q$ remains conserved during its motion.
Then the simplest pair of the time reversed paths is: $q_-(t')=
2R\sin (2\pi t'/t)$ and $q_+(t')$ is an even function of $t'$. Substituting
these paths into (\ref{SR0},\ref{SI0}) after a simple integration we find
\begin{equation}
S_R=0,\;\;\;\;\;S_I(t)=\frac{2R^2C^2}{\omega}\coth
\left( \frac{\omega}{2T}\right)
\left(\frac{t}{2\pi }\right)^2
\frac{\sin ^2(\omega t/2)}{(1-\omega^2t^2/4\pi^2)^2},
\label{SRI1}
\end{equation}
The result (\ref{SRI1}) demonstrates that the return probability $W_2$
for a particle $q$ after a time $t$ for the time reversed paths
acquires the factor
\begin{equation}
F=\exp (-S_I(t)),
\label{FSI}
\end{equation}
due to interaction with a harmonic oscillator with a frequency $\omega$.
The probability $W_1$ remains unaffected.

The whole consideration can be trivially generalized to the case
of more complicated time reversed paths with an arbitrary winding
numbers $m$. In this case we again find an additional factor (\ref{FSI})
in the interference term $W_2$ and $S_I$ is determined
by an expression similar to (\ref{SRI1}) which also depends on $m$. The
term $S_R$ is again zero for such paths for all $m$. Since no new effects
emerge at $m>1$, in what follows we will analyze only the simplest
case $m=1$ (\ref{SRI1}).

The first conclusion one can draw from the result (\ref{SRI1}) is that
no qualitative changes in the system behavior emerges if one varies
the temperature $T$. The value $S_I(t)$
is smaller at low $T \ll \omega$ as compared to the high temperature limit
$T \gg \omega$ but the effect persists even at $T=0$. Thus the relation
between $T$ and $\omega$ turns out to be important only in a
quantitative sense, no qualitative dependence on
this relation should be expected.

A much more important parameter is $\omega t$. We see that
the value $S_I(t)$ and hence $W_2$ oscillate in time with a period
$2\pi/\omega$. After each such period the interference term $W_2$
restores its ``nonineracting'' value while at all intermediate times
the value $W_2$ is smaller than in the noninteracting case. In this
situation we still cannot speak about decoherence: the system keeps
information about its initial phase and periodically
returns to its initial state. On top of that if the radius of the
ring $R$ is constant in time (see below) we have $S_I \propto 1/t^2$,
i.e. the oscillations of $W_2$ practically disappear in the long
time limit. All these results are not surprizing: one should not
expect to find the decoherence effect in the system of two quantum
mechanical particles.

Quantum decoherence appears after the next step:
we again modify our model coming from the interaction with one oscillator
to the infinite set of oscillators with the ohmic spectrum (\ref{ohm}). In this
case after the integration of (\ref{SRI1}) over $\omega$ one gets
\begin{equation}
S_I=\frac{\eta R^2}{\pi}\int zdz\frac{\sin^2(\pi z)}{(1-z^2)^2}
\coth \left(\frac{\pi z}{Tt}\right) .
\label{SI5}
\end{equation}
This equation yields $S_I \sim \eta R^2$ at $T=0$ and
$S_I \sim \eta TR^2t$ at $Tt \gg 1$.
These results are in a nice qualitative agreement with those obtained
in the exactly solvable model studied above. There the exponential
decay of the off-diagonal elements of the initial
density matrix $\rho (q_{-i})\propto \exp (-t/\tau_0)$ with
$\tau_0 \sim 1/\eta Tq_{-i}^2$ was found at any $T>0$ while
at $T=0$ a power law decay (\ref{power}) was observed. In both cases
the similarity is obvious if one interchanges $R$ and $q_{-i}$.

We observe from (\ref{FSI}) that if $S_I$ is small the influence
functional $F \simeq 1$ and no decoherence occurs. However for
$S_I \gg 1$ the interference of the time-reversed paths is completely
suppressed, and quantum dephasing takes place. By setting $S_I \sim 1$ we can,
therefore, define the typical size of the ring $R \sim L_{\varphi}$ beyond
which the effect of quantum decoherence becomes important. In our
particular example the characteristic dephasing length $L_{\varphi}$
decreases as $1/t$ at $Tt \gg 1$ and it is constant
\begin{equation}
L_{\varphi} \sim 1/\sqrt{\eta }
\label{L}
\end{equation}
at $T=0$. This is the effect of quantum decoherence.

Note that by fixing $R$ and increasing $t$ in our problem
we effectively decrease the particle
velocity $v_0=2\pi R/t$ which is eventually sent to zero as
$t$ approaches infinity.
Therefore it is quite natural that at $T=0$ the interference term $W_2$,
although suppressed by a factor $F=\exp (-S_I)$, does not decay in
time. Qualitatively the same property is observed in our result for the
decoherence time in a disordered metal (\ref{1}): if we treat $v_F$ as a
formal parameter which may be put equal to zero we will immediately
arrive at a zero decoherence rate $1/\tau_{\varphi}$ in this limit. The
same is true for the high temperature result \cite{AAK,CS,SAI,Imry}.

The above situation is, however, not very relevant for a metal where
conducting electrons move with the velocity $v_0 \simeq v_F$ which
absolute value does not change in time. Thus to account for that we
should rather keep the velocity of a particle $v_0$ fixed.
This implies
that as we increase $t$ the radius $R$ for a classical return path
increases linearly with time. In other words, we can slightly modify
our model allowing our particle to choose the ring with a proper $R$
for each time $t$ (see Fig. 1c). Then we immediately observe that
$S_I$ (\ref{SI5}) grows in time as $S_I \propto t^2$ at $T=0$ and
$S_I \propto t^3$ at $Tt \gg 1$, i.e. in this case the decay of
the quantum interference term $W_2$ caused by interaction with
the Caldeira-Leggett bath of oscillators is {\it faster than exponential}.
Finally, let us note that in a disordered metal, although the
length of the electron trajectory increases linearly with time,
the dynamics is diffusive and the effective loop size grows as
$R^2 \sim Dt$. Substituting this expression into (\ref{SI5}) one obtains
\begin{equation}
F=
\left\{\begin{array}{ll}
\exp (-t/\tau_{\varphi}),  \;\;\;\;\; \tau_{\varphi}=a/D\eta, & Tt \ll 1\\
\exp (-(t/\tau_{\varphi})^2),  \;\;\;\; \tau_{\varphi}=b(D\eta T)^{-1/2},
& Tt \gg 1
\end{array}\right.
\label{FT}
\end{equation}
where $a$ and $b$ are unimportant numerical
coefficients of order one.

Although the above simple model cannot be directly applied to
disordered metals it demonstrates several important
properties which will be also observed in a rigorous calculation.
One such property is that the suppression of quantum interference
between time reversed paths increases
if the size of the loop grows in time. Another property is that
oscillators with $\omega \gg T$ may give the dominating contribution
to dephasing. In the above model the frequencies $\omega \sim 1/t$
give the maximum contribution, but if the spectrum of the
problem is different (as e.g. in a disordered metal, see below)
high frequency modes $\sim \omega_c$ may also become important.
The physical reasons for this
conclusion were already clarified above: in order to
have energy exchange in an interacting system
it is not necessary to excite the high frequency oscillators,
broadening of their ground state levels is sufficient. It is well
known that the high frequency cutoff $\omega_c$
enters the expression for the interaction induced decoherence
rate of a quantum particle in the periodic \cite{SZ,Schmid2}
and the double well potentials \cite{ChLeg}. The same is
observed here for an exactly solvable model of a free damped particle
in the limit $T=0$.

Our model also demonstrates at which step of our
calculation the effect of dephasing appears. Interaction of
the particle $q$ with one harmonic oscillator leads to the
oscillations of the interference term $W_2$ with the oscillator frequency
$\omega$. These oscillations are natural since the initial state
is not the eigenstate of the system. Yet no dephasing appears.
If coupling to {\it many} oscillators with different frequencies
is introduced the probability $W_2$ will be {\it always} suppressed
and the system will {\it never} return to its initial state.
Obviously this mechanism of dephasing has nothing to do with the temperature
of the environment and it persists even at $T=0$. In this limit the exponential
decay of $W_2$ in time is due to increase of the loop size with $t$.
The phase breaking length $L_{\varphi}$ (\ref{L})
depends only on the interaction strength and appears to be (roughly)
insensitive to the particular (e.g. ballistic or diffusive) type
of a particle motion.

\subsection{Scattering problem and perturbation theory}

One might wonder what is the relation between the above analysis
and the standard perturbative treatment of a scattering problem
for a quantum particle interacting with a harmonic oscillator.
Beside its general importance the scattering approach is
also of a practical relevance because it is frequently applied
to conductance calculations in mesoscopic systems.

The usual definition of a scattering problem operates with in-
and out-scattering states measured after exactly {\it infinite} time.
In other words, the limit $t \to \infty$ is taken already at
the very first step, and the whole calculation is carried out only
in this limit. This is sufficient in many physical situations, but
not for our problem due to the reasons to be clarified below.
Here we will keep $t$ finite (although possibly large) throughout
the calculation and let it go to infinity in the end. With this
in mind a direct connection to the scattering problem
can be easily established.

Consider a quantum particle $q$ scattered on
a harmonic oscillator with a frequency $\omega$. Before scattering
the oscillator is assumed to be in its ground state, and the particle
is in the state with wave function $\psi_i$ and the energy $E_i$.
Assuming the interaction to be of the same form as in (\ref{SqQ1})
and proceeding perturbatively in the interaction strength $C$ one
can easily derive the probability $W(t)$ for a particle $q$ to leave the
state $\psi_i$ after the time $t$. It is just the sum of the
transition probabilities $W_{if}$ into all possible final states
$\psi_f$ which are orthogonal to $\psi_i$ in the absence of interaction.
One easily finds (see e.g. \cite{FH})
\begin{equation}
W=\sum_{f\neq i}W_{if}=2\sum_{f\neq i}|\langle i|Cq|f\rangle |^2
\frac{\sin^2(\Delta E_{if}t/2)}{\omega (\Delta E_{if})^2},
\label{GR}
\end{equation}
where $\Delta E_{if}=\omega +E_f-E_i$. This result implies
$dW/dt \propto \sin (\Delta E_{if}t)/\Delta E_{if}$, i.e.
at large $t$ the transition rate experiences fast oscillations
and approaches $\delta (\Delta E_{if})$ at $t \to \infty$.
If $E_f>E_i$ for all $f$ (i.e. $\psi_i$ describes the ground state of
the noninteracting problem) the transitions are highly improbable
at sufficiently long times and one may conclude that with the
dominating probability the particle
remains in its initial state and no quantum dephasing takes place.
We would like to emphasize, however, that the time average of the
escape probability $W$ (\ref{GR}) is not equal to zero already in this
case. This is a direct consequence of the energy-time uncertainty
principle.

Let us now consider scattering on {\it many} oscillators.
As before we will assume that the frequency spectrum
for these oscillators is ohmic (\ref{ohm}). In order to define
the scattering problem we also assume that interaction exists
only in a certain space region $-q_0/2<q<q_0/2$, i.e. we put
$C=C_0(\Theta (q+q_0/2)-\Theta (q-q_0/2))$. Without interaction
the eigenstates of a problem are the plane waves
$\psi_{i/f}(q,t)=(L)^{-1/2}\exp (ip_{i/f}q-iE_{i/f}t)$ where
$E_{i/f}=p_{i/f}^2/2m$.
The transition matrix elements can be easily evaluated. They give
an important contribution for $|p_f-p_i|q_0<1$ in which
case we get $\langle i|Cq|f\rangle |^2 \sim C_0^2q_0^6(p_f-p_i)^2/L^2$.
Subsituting this expression into (\ref{GR})
and carrying out the summation over the final states $\sum_f \to L\int dp_f$
and over the oscillator frequencies (making use of (\ref{subs})
with $C \to C_0$), we find
\begin{equation}
W(t) \sim \frac{\eta q_0^6}{L}\int_0^{\omega_c} \omega d\omega
\int_{-1/q_0}^{1/q_0}dp_f (p_f-p_i)^2
\frac{\sin^2(\Delta E_{if}t/2)}{(\Delta E_{if})^2}.
\label{GR2}
\end{equation}
Further assuming that the initial particle energy is small $E_i \to 0$
after simple integrations we obtain
\begin{equation}
W(t) \sim \frac{\eta q_0^3}{L}\ln (\omega_c \mbox{min}(t,mq_0^2)).
\label{GR3}
\end{equation}
We observe that the escape probability $W(t)$ is not zero and, moreover,
it is not necessarily small. The reason for that is transparent.
Although the contribution of each oscillator to $W(t)$ is small, it
is not zero at any finite $t$ due to the energy-time uncertainty principle.
The sum of these small contributions from {\it many} oscillators is finite
and yields the result (\ref{GR3}). We would like to emphasize that by no
means this result is in contradiction with the energy conservation law, rather
it demonstrates that one should be careful applying the energy
arguments to describe the time evolution of an interacting quantum
system, especially if it consists of an infinite number of degrees of
freedom. Even at times much larger than the characteristic
time scale (in our problem the relevant time scale is set by the
dephasing time $\tau_{\varphi}$) the energy uncertainty may be sufficient
for $W(t)$ to significantly differ from zero.

In order to observe the relation of a perturbative expression (\ref{GR3})
to our previous results we set $q_0$ to be of order of
the system size $q_0 \sim L$. After that the similarity between (\ref{GR3})
and e.g. eqs. (\ref{power},\ref{SI5}) becomes completely obvious. Requiring
that $W(t,L) \sim 1$ (the escape is complete) and using (\ref{GR3})
with $q_0 \sim L$ one immediately arrives at the estimate (\ref{L}) for the
decoherence length $L_{\varphi} \sim 1/\sqrt{\eta}$ derived previously
with the quasiclassical analysis of the exact influence functional.
This result leaves no room for doubts concerning the validity of the
quasiclassical description of quantum dephasing at $T=0$. In fact,
the comparison of the results (\ref{power},\ref{GR3}) with (\ref{SI5})
demonstrates that the quasiclassical approximation rather
{\it underestimates} the dephasing effect of interaction: $L_{\varphi}$
may only become shorter if fluctuations around the classical trajectory
are taken into account.

Another obvious conclusion is that at any $T$ (including $T=0$) the effect
of quantum decoherence in not only due to low frequency oscillators which,
moreover, can be even completely unimportant. The relative contribution
of oscillators with small $\omega$ depends on the particular form of
the spectrum ($\sim \omega^{\gamma}$),
being more important in the subohmic case $\gamma <1$ and practically
irrelevant in the superohmic case $\gamma >1$ when the high frequency
oscillators yield the main effect. The latter situation will be
encountered in the next section where it will be shown that
for a $d$-dimensional disordered metal one has $\gamma =1+d/2$.
Again we emphasize that the above results {\it do not} imply
that the processes with high energy transfers are important for
dephasing at {\it any} $T$. It is {\it erroneous} to interpret
the parameter $\omega$ as describing the energy transfer: in our
calculation the integral over $\omega$ always represents the summation
over the bath oscillators (cf. eqs. (\ref{ohm}-\ref{subs})).
No high frequency oscillators need to
be excited, (small) energy uncertainty for {\it infinitely} many
oscillators is sufficient to provide (large) dephasing for a particle $q$.

\section{Quantum Decoherence in a Disordered Metal}

In order to provide a quantitative description of the effect of
quantum decoherence in disordered metals it is necessary to go beyond
the simple model considered in the previous section and
account for $(a)$ Fermi statistics and the Pauli
principle, $(b)$ the specifics of Coulomb interaction in a $d$-dimensional
system and $(c)$ the effect of disorder. The corresponding analysis is
presented below.

\subsection{Density matrix and effective action}

Our starting point is
the standard Hamiltonian for electrons in a disordered metal
$H_{\rm el}=H_0+H_{\rm int}$, where
\begin{equation}
H_0= \int {\rm d}{\bf r}\psi^+_{\sigma}({\bf r})
\left[-\frac{\nabla^2}{2m}-\mu +U({\bf r})\right]\psi_\sigma ({\bf r}),
\label{H0}
\end{equation}
\begin{equation}
H_{int}=\frac{1}{2}
\int {\rm d}{\bf r}\int {\rm d}{\bf r'}\psi^+_{\sigma}({\bf r})\psi^+_{\sigma'}({\bf r'})
e^2v({\bf r}-{\bf r'})\psi_{\sigma'}({\bf r'})\psi_{\sigma}({\bf r}).
\label{Hint}
\end{equation}
Here $\mu$ is the chemical potential, $U({\bf r})$ accounts for a random
potential due to nonmagnetic impurities, and
$v({\bf r})=1/|{\bf r}|$ represents
the Coulomb interaction between electrons.

Let us define the electron Green-Keldysh function \cite{Keldysh}
\begin{equation}
\hat G\equiv \left( \begin{array}{cc}  G_{11} & - G_{12} \\
G_{21} & - G_{22} \end{array} \right)=
\frac{\int {\rm D}V_1 {\rm D}V_2
\; \hat G_V\; e^{{\rm i}S[V_1,V_2]}}
{\int {\rm D}V_1{\rm D}V_2\; e^{{\rm i}S[V_1,V_2]}},
\label{G1}
\end{equation}
where
\begin{equation}
{\rm i}S[V_1,V_2]=2{\rm Tr}\ln\hat G_V^{-1}+ {\rm i}
\int\limits_0^t {\rm d}t'\int
{\rm d}{\bf r} \frac{(\nabla V_1)^2-(\nabla V_2)^2}{8\pi}.
\label{lndet}
\end{equation}
Here we performed a standard
Hubbard-Stratonovich transformation introducing the path integral
over a scalar potential field $V$ in order to decouple the $\psi^4$-interaction
in (\ref{Hint}). In (\ref{G1},\ref{lndet}) we explicitely defined the
fields $V_1(t)$ and $V_2(t)$ equal to $V(t)$ respectively on the forward
and backward parts on the Keldysh contour \cite{Keldysh}.

The matrix function $\hat G_V$ obeys the equation
\begin{equation}
\left({\rm i}\frac{\partial}{\partial t_1}-\hat
H_0({\bf r}_1)+e\hat V(t_1,{\bf r}_1)\right) \hat G_V=\delta(t_1-t_2)
\delta({\bf r}_1-{\bf r}_2);
\label{Schrodinger}
\end{equation}
where
\begin{equation}
\hat H_0=H_0\hat 1=\left( \begin{array}{c}
                \frac{-\nabla^2}{2m}-\mu+U({\bf r}) \qquad 0 \\
                0 \qquad \frac{-\nabla^2}{2m}-\mu+U({\bf r})
                \end{array} \right);\;\;\;\;\;\;\;
\hat V=\left( \begin{array}{cc}
                V_1(t,{\bf r}) & 0 \\
                0    & V_2(t,{\bf r})
                \end{array} \right).
\label{V}
\end{equation}

The solution of (\ref{Schrodinger}) is fixed by the Dyson equation
\begin{equation}
\hat G_V(t_1,t_2)=\hat G_0(t_1,t_2) -\int\limits_0^t {\rm d}t'
\hat G_0(t_1,t') e\hat V(t')\hat G_V(t',t_2).
\label{Dyson}
\end{equation}
The matrix $\hat G_0$ is the electron Green-Keldysh function
without the field $V$.

It is well known that the 1,2-component of the Green-Keldysh
matrix ${\bf \hat G}$
is directly related to the exact electron density matrix
\begin{equation}
\rho (t; {\bf r},{\bf r'})=-{\rm i}G_{12}(t,t;{\bf r},{\bf r'})
=\langle \rho_V (t; {\bf r},{\bf r'}\rangle_{V_1,V_2},
\label{rhoG}
\end{equation}
where we also defined the ``density
matrix'' $\rho_V(t)$ related to the 1,2-component
of the matrix $\hat G_V$ and performed the average over the
fields $V_1$ and $V_2$ as defined in (\ref{G1}).
The density matrix $\rho$ contains all necessary information about
the system dynamics in the presence of interaction.

Making use of eqs. (\ref{Schrodinger}-\ref{Dyson}) after some formal
manipulations (see \cite{GZ} for details) one arrives at the
equation describing the time evolution of the density matrix:
\begin{equation}
{\rm i}\frac{\partial\rho_V}{\partial t}=
[H_0-eV^+,\rho_V] - (1-\rho_V)\frac{eV^-}{2}\rho_V -
\rho_V\frac{eV^-}{2}(1-\rho_V),
\label{rho5}
\end{equation}
where we defined $V^+=(V_1+V_2)/2$ and $V^-=V_1-V_2$.
It is important to emphasize that the derivation of this equation
was performed {\it without any approximation}, i.e. the result
(\ref{rho5}) {\it is exact}. The equation (\ref{rho5}) fully
accounts for the Pauli principle which is important for
the fluctuations of the field $V^-$. This field is irrelevant for dephasing.
It is quite obvious from (\ref{rho5}) that the field
$V^+(t,{\bf r})$ plays the same role as an external field. All electrons
move collectively in this field, its presence is equivalent to local
fluctuations of the Fermi energy $\mu \to \mu +eV^+(t,{\bf r})$.
The Pauli principle does not play any role here. There is no way how
the density matrix $\rho$ can ``distinguish'' the {\it intrinsic}
fluctuating field $V^+$ from the stochastic
{\it external} field, be it classical or quantum. Since the external field is
known to lead to dephasing of the wave function, the field $V^+$ should
produce the same effect. As the equation (\ref{rho5}) is exact this
conclusion is general and {\it does not depend on approximations}.

In order to proceed further let us assume Coulomb interaction
to be sufficiently weak and expand the action (\ref{lndet}) in
powers of $V$ up to terms proportional to $V^2$. After a
straightforward calculation (see e.g. \cite{GZ}) one finds
\begin{eqnarray}
{\rm i}S[V_1,V_2]=
{\rm i}\int\frac{{\rm d}\omega {\rm d}^3k}{(2\pi)^4}
V^-(-\omega,-k)\frac{k^2\epsilon(\omega,k)}{4\pi}V^+(\omega,k)-
\nonumber \\
-\frac{1}{2}
\int\frac{{\rm d}\omega {\rm d}^3k}{(2\pi)^4}
V^-(-\omega,-k)\frac{k^2{\rm Im}\epsilon(\omega,k)}{4\pi}
\coth\left(\frac{\omega}{2T}\right)
V^-(\omega,k);
\label{actionf}
\end{eqnarray}
where $\epsilon (\omega , k)$ is the dielectric susceptibility of
a disordered metal
\begin{equation}
\epsilon(\omega,k)=1+\frac{4\pi\sigma}{-{\rm i}\omega +Dk^2}.
\label{eps}
\end{equation}
Here $\sigma =2e^2N_0D$ is the classical Drude conductivity, $N_0$ is the
metallic density of states and $D=v_Fl/3$ is the diffusion coefficient.
For the sake of simplicity in eq. (\ref{eps}) we disregarded the
phonon contribution which will not be important for us here.
The expression (\ref{eps}) is valid for small wave vectors $k < 1/l$ and
small frequencies $\omega < 1/\tau_e=v_F/l$.

Note, that if one considers only nearly uniform in
space ($k \approx 0$) fluctuations of the field $V$ one immediately
observes that eqs. (\ref{actionf},\ref{eps}) {\it exactly} coincide with
the real time version of the Caldeira-Leggett action
\cite{CL,Schmid1,SZ} in this limit (cf. (\ref{F2}-\ref{SI0},\ref{subs})).
Taking into account only uniform fluctuations of the
electric field one can also derive the Caldeira-Leggett
action expressed in terms of the {\it electron coordinate only}.
In this
case the effective viscosity $\eta$ in the Caldeira-Leggett influence
functional is $\eta \sim e^2R_s$, where $R_s$ is the sample resistance
(in contrast to the
effective viscosity for the field $V$ which is proportional to $1/R_s$).

For our present purposes it is not sufficient to
restrict ourselves to uniform fluctuations of the collective
coordinate $V$ of the electron environment.
The task at hand is to evaluate the kernel of the operator
$$
J= \sum_V U|V\rangle\langle V|U^+
$$
where the sum runs over the states of the
electromagnetic environment with all possible $k$ and $\omega$.
Averaging over $V^+,V^-$ amounts to calculating
Gaussian path integrals with the action (\ref{actionf})
and can be easily performed. We obtain
$$
J(t,t';{\bf r}_{1f},{\bf r}_{2f};{\bf r}_{1i},{\bf r}_{2i})=
\int\limits_{{\bf r}_1(t')={\bf r}_{1i}}^{{\bf r}_1(t)={\bf r}_{1f}}
{\rm D}{\bf r}_1\int\limits_{{\bf r}_2(t')={\bf r}_{2i}}^{{\bf r}_2(t)
={\bf r}_{2f}}{\rm D}{\bf r}_2\int {\rm D}{\bf p}_1 {\rm D}{\bf p}_2\times
$$
\begin{equation}
\times \exp\big\{{\rm i}S_0[{\bf r}_1,{\bf p}_1]-
{\rm i}S_0[{\bf r}_2,{\bf p}_2]-
{\rm i}S_R[{\bf r}_1,{\bf p}_1,{\bf r}_2,{\bf p}_2]-
S_I[{\bf r}_1,{\bf r}_2]\big\};
\label{JS}
\end{equation}
where
\begin{equation}
S_0[{\bf r},{\bf p}]=\int\limits_{t'}^t {\rm d}t''
\bigg({\bf p\dot r} - \frac{{\bf p}^2}{2m} - U({\bf r})\bigg)
\label{S_0}
\end{equation}
is the electron action,
\begin{eqnarray}
S_R[{\bf r}_1,{\bf p}_1,{\bf r}_2,{\bf p}_2]&=&
\frac{e^2}{2}\int\limits_{t'}^t {\rm d}t_1 \int\limits_{t'}^t {\rm d}t_2
\big\{R(t_1-t_2,{\bf r}_1(t_1)-{\bf r}_1(t_2))
\big[1-2n\big({\bf p}_1(t_2),{\bf r}_1(t_2)\big)\big]-
\nonumber \\
&&
-R(t_1-t_2,{\bf r}_2(t_1)-{\bf r}_2(t_2))
\big[1-2n\big({\bf p}_2(t_2),{\bf r}_2(t_2)\big)\big]
\nonumber \\
&&
+R(t_1-t_2,{\bf r}_1(t_1)-{\bf r}_2(t_2))
\big[1-2n\big({\bf p}_2(t_2),{\bf r}_2(t_2)\big)\big]-
\nonumber \\
&&
-R(t_1-t_2,{\bf r}_2(t_1)-{\bf r}_1(t_2))
\big[1-2n\big({\bf p}_1(t_2),{\bf r}_1(t_2)\big)\big]
\big\},
\label{SR}
\end{eqnarray}
where $n ({\bf p},{\bf r})$ is the occupation number and
\begin{eqnarray}
S_I[{\bf r}_1,{\bf r}_2]&=&
\frac{e^2}{2}\int\limits_{t'}^t {\rm d}t_1 \int\limits_{t'}^t {\rm d}t_2
\bigg\{I(t_1-t_2,{\bf r}_1(t_1)-{\bf r}_1(t_2))+
I(t_1-t_2,{\bf r}_2(t_1)-{\bf r}_2(t_2))-
\nonumber \\
&&
-I(t_1-t_2,{\bf r}_1(t_1)-{\bf r}_2(t_2))-
I(t_1-t_2,{\bf r}_2(t_1)-{\bf r}_1(t_2))\bigg\}.
\label{SI}
\end{eqnarray}
In equilibrim $n$ is just the Fermi function. In this case
at the scales $|{\bf r}|> l$
the functions $R$ and $I$ are defined by the equations
\begin{eqnarray}
R(t,{\bf r})&=&\int\frac{{\rm d}\omega {\rm d}^3k}{(2\pi)^4}\medskip
\frac{4\pi}{k^2\epsilon(\omega,k)}e^{-i\omega t+i{\bf kr}}
\label{R}\\
I(t,{\bf r})&=&\int\frac{{\rm d}\omega {\rm d}^3k}{(2\pi)^4}\medskip
{\rm Im}\left(\frac{-4\pi}{k^2\epsilon(\omega,k)}\right)
\coth\bigg(\frac{\omega}{2T}\bigg)
e^{-i\omega t+i{\bf kr}} .
\label{RI}
\end{eqnarray}

The expression in the exponent of eq. (\ref{JS}) defines
the real time effective action of the electron propagating in a
disordered metal and interacting with other electrons. The first
two terms represent the electron action $S_0$ (\ref{S_0}) on
two branches of the Keldysh contour
while the last two terms $S_R$ and $S_I$ determine the influence
functional (cf. eqs. (\ref{J},\ref{F2}-\ref{SI0},\ref{subs}))
of the effective electron
environment. As can be seen from eqs. (\ref{SR}-\ref{RI}) this influence
functional is not identical to one derived in the
Caldeira-Leggett model. However on a qualitative level the similarity
is obvious: in both models the influence functionals describe the effect of a
certain effective {\it dissipative} environment.

\subsection{Decoherence time}

Let us first neglect the terms $S_R$ and $S_I$ describing the effect
of Coulomb interaction. Then in the quasiclassical limit $p_Fl \gg 1$
the path integral (\ref{JS}) is dominated by the saddle point trajectories for
the action $S_0$:
\begin{equation}
{\bf \dot p}=-\partial H_0/\partial {\bf r},
\quad {\bf \dot r}=\partial H_0/\partial {\bf p}
\label{Hamilton}
\end{equation}
with obvious boundary conditions ${\bf r}_1(t')={\bf r}_{1i}$,
${\bf r}(t)={\bf r}_{1f}$ for the action $S_0[{\bf r}_1,{\bf p}_1]$ and
${\bf r}_2(t')={\bf r}_{2i}$,
${\bf r}_2(t)={\bf r}_{2f}$ for the action $S_0[{\bf r}_2,{\bf p}_2]$.

Since in a random potential $U({\bf r})$ there is in general no
correlation between different classical paths ${\bf r}_1(t)$ and
${\bf r}_2(t)$ these paths give no contribution to the
integral (\ref{JS}): the difference of two actions $S_0$ in the exponent
 may have an arbitrary value and the result
averages out after summation. Thus only the paths with
$S_0[{\bf r}_1,{\bf p}_1]\simeq S_0[{\bf r}_2,{\bf p}_2]$ provide
a nonzero contribution to the path integral (\ref{JS}).
Two different classes
of paths can be distinguished (see e.g. \cite{CS}):

i) The two classical paths are almost the same:
${\bf r}_1(t'')\simeq {\bf r}_2(t'')$, ${\bf p}_1(t'')\simeq {\bf p}_2(t'')$
(cf. Fig. 1a).
For such pairs we obviously have ${\bf r}_{1i}\simeq {\bf r}_{2i}$ and
${\bf r}_{1f}\simeq {\bf r}_{2f}$. Physically this corresponds
to the picture of electrons propagating as nearly classical particles.
In the diffusive limit these paths give rize to diffusons (see e.g.
\cite{AAK}) and yield the standard Drude conductance.

ii) The pairs of time reversed paths. In this case
${\bf r}_{1i}\simeq {\bf r}_{2f}$, ${\bf r}_{1f}\simeq {\bf r}_{2i}$
(cf. Fig. 1b,c). In the path integral (\ref{J}) the trajectories ${\bf r}_1$
and ${\bf r}_2$ are related as ${\bf r}_2(t'')\simeq {\bf r}_1(t+t'-t'')$
and ${\bf p}_2(t'')\simeq -{\bf p}_1(t+t'-t'')$. In a disordered
metal these paths correspond to Cooperons and give rize
to the weak localization correction to conductivity $\delta \sigma_d$.
This correction is expressed in terms of the time integrated
probability $W(t)$  for all diffusive paths to return to the same
point after the time $t$ (see e.g. \cite{AAK,CS}).
In the absence of any kind of interaction
which breaks the time reversal symmetry this value coincides with the classical
return probability and is given by the formula
$W_0(t)=(4\pi Dt)^{-d/2}a^{-(3-d)}$, where $d$ is the system dimension
and $a$ is the transversal sample size.

The weak localization correction $\delta\sigma_d$ diverges for $d \leq 2$.
This divergence can be cured by
introducing the upper limit cutoff
at a certain time $\tau_\varphi$. This time is usually
reffered to as decoherence time. One finds \cite{AAK,CS}:
\begin{equation}
\delta\sigma_d=\left\{\begin{array}{ll}
     -\frac{e^2}{2\pi^2 }\ln\left(\frac{\tau_\varphi}{\tau_e}\right), & d=2 \\
    -\frac{e^2}{\pi }\sqrt{D\tau_\varphi}, & d=1.
     \end{array}\right.
\label{dsigma1}
\end{equation}

The physical reason for the existence of a finite $\tau_{\varphi}$
is the electron-electron interaction which breaks the time reversal symmetry
in our problem.
To evaluate $\tau_\varphi$ we first note that the functions $R$ and $I$
(\ref{RI}) change slowly at distances of the order of the Fermi
wavelength $1/p_F$. Therefore we may put ${\bf r}_1(t'')={\bf r}(t'')$,
${\bf r}_2(t'')={\bf r}_t(t'')\equiv{\bf r}(t+t'-t'')$.
Here ${\bf r}(t'')$ is a classical trajectory with the initial point
${\bf r}(t')=0$ and the final
point $|{\bf r}(t)|< l$, i.e. we consider trajectories
which return to the vicinity of the initial point.
The contribution from the time reversed paths $W_2$ to the
return probability has the form
\begin{equation}
W_2(t-t')\simeq W_{20}(t-t')F, \;\;\;
F=\left\langle e^{-{\rm i}S_R[t,t';{\bf r},{\bf p};{\bf r}_t,{\bf p}_t]
-S_I[t,t';{\bf r},{\bf r}_t]}
\right\rangle_{r},
\label{W}
\end{equation}
where $W_{20}$ is the probability without interaction and the average
is taken over all diffusive paths returning to the initial point.
The value $F$ in (\ref{W}) decays exponentially in time, therefore we
may put the average inside the exponent. It is easy to observe that
the term $S_R$ gives no contribution to this average. Working out
the average of $S_I$ we obtain
\begin{equation}
F=e^{-\left\langle S_I[t,0;{\bf r},{\bf r}_t]
\right\rangle_{r}} = \exp \left(-t
e^2\int\limits_{-\infty}^{+\infty} {\rm d}t
\langle I(t,{\bf r}(t)-{\bf r}(0))\rangle_{r}\right)
\label{tau}
\end{equation}
To find the average over the diffusive paths, we introduce the Fourier
transform of the function $I(t,{\bf r})$ and replace
$\langle e^{-ik({\bf r}(t)-{\bf r}(t'))}\rangle_{r}$ by $e^{-Dk^2|t-t'|}$.
Then we get
\begin{equation}
\ln F(t)=
-\frac{te^2}{a^{3-d}}\int\frac{{\rm d}\omega {\rm d}^dk}{(2\pi)^{d+1}}
{\rm Im}\left(\frac{-4\pi}{k^2\epsilon(\omega,k)}\right)
\coth\left(\frac{\omega}{2T}\right)
\frac{Dk^2}{\omega^2+D^2k^4}
\label{3d}
\end{equation}

Let us first consider a quasi-one-dimensional system with $a \leq l$.
Making use of eqs. (\ref{3d},\ref{eps}) and integrating over $k$ we find
\begin{equation}
\ln F(t) =-t\frac{e^2\sqrt{2D}}{\sigma_1 }
\int\limits_{1/t}^{1/\tau_{e}}
\frac{{\rm d}\omega}{2\pi}\frac{\coth (\omega /2T)}{\sqrt{\omega}}.
\label{tau1}
\end{equation}
The upper cutoff in (\ref{tau1}) is chosen
at the scale $\sim 1/\tau_e$ because at higher $\omega$ the
diffusion approximation becomes incorrect. From (\ref{tau1})
we obtain
\begin{equation}
F=
\left\{\begin{array}{ll}
\exp (-t/\tau_{\varphi}),  \;\;\;\;\; \tau_{\varphi}\simeq
\pi (3/2)^{1/2}\sigma_1/e^2v_F, & t\ll 1/T^2\tau_e\\
\exp (-(t/\tau_{\varphi})^{3/2}),  \;\;\;\; \tau_{\varphi}\simeq
0.5(\pi\sigma_1 /D^{1/2}T)^{2/3},
&t\gg 1/T^2\tau_e
\end{array}\right.
\label{F1D}
\end{equation}
In the long time limit (which cannot be reached at
$T \to 0$) the decay of $F$ is faster than
exponential (cf. \cite{Imry}). At sufficiently low
$T <(\tau_e\tau_{\varphi})^{-1/2}$ this difference becomes important
only at $t \gg \tau_{\varphi}$ where $F$ is already exponentially
suppressed. For smaller $t$ the decay is exponential and the decoherence
rate increases linearly with $T$. Defining the effective decoherence
length $L_{\varphi} \simeq \sqrt{D\tau_{\varphi}}$, in the temperature
interval
$1/2\tau_\varphi(0) < T <(\tau_e\tau_{\varphi}(0))^{-1/2}$ one finds
\begin{equation}
\frac{1}{L_\varphi^2(T)}=\frac{1}{L_\varphi^2(0)}+
2\sqrt{2}\frac{L_{\varphi}(0)e^2}{\pi D\sigma_1}T,
\label{length}
\end{equation}
where $L_{\varphi}(0)$ is the dephasing length at $T=0$.
At temperatures lower than $1/2\tau_\varphi(0)$ the decoherence
time and length are almost temperature independent.
With the aid of eqs. (\ref{dsigma1},\ref{F1D}) it is also
easy to find the weak localization
correction $\delta \sigma_1$ to the Drude conductance. In the limit $T=0$
we obtain
\begin{equation}
\frac{\delta \sigma_1}{\sigma_1}=-\frac{e^2}{\pi \sigma_1
}\sqrt{D\tau_{\varphi}} \approx - \frac1{p_Fs^{1/2}}, \label{delsig}
\end{equation}
i.e. $\delta \sigma_1 \approx - \sigma_1 /\sqrt{N_{\rm ch}}$,
where $N_{\rm ch} \sim p_F^2s$ is
the effective number of conducting channels in a 1d mesoscopic system and
$\sigma_d = \sigma a^{3-d}$ is the Drude conductance of a
$d$-dimensional sample.
Thus for $N_{\rm ch} \gg 1$ the weak localization correction
is {\it parametrically} smaller than the Drude conductance
even at $T=0$.

For 2d and 3d systems the same analysis yields
\begin{eqnarray}
\frac{1}{\tau_\varphi} & = & \frac{e^2}{4\pi\sigma_2 \tau_e}
[1+2T\tau_e\ln(T\tau_\varphi)],  \; \; \; \; \; \; \;  \quad{\rm 2d},
\nonumber \\
\frac{1}{\tau_\varphi} & = & \frac{e^2}
{3\pi^2\sigma\sqrt{2D}\tau_e^{3/2}}[1+6(T\tau_e)^{3/2}],\;  \quad{\rm 3d},
\label{11}
\end{eqnarray}
The influence functional $F$ decays exponentially $F=\exp (-t/\tau_{\varphi})$
except for the case of 2d systems at high $T$ where
one has $\ln F \propto -t\ln t$. We observe that also in this case
the difference from a purely exponential decay is not significant and
can be ignored.

\subsection{Further remarks}

The above formalism provides a rigorous description of the
effect of quantum decoherence in disordered metals and allows
to derive the decoherence rate $1/\tau_{\varphi}$ at low $T$
for various dimensions. The physical reasons for dephasing of
electron wave functions remain the same as in the case of a particle
interacting with the Caldeira-Leggett bath of oscillators. The
corresponding discussion is presented in the previous section.
Here we only add several comments.

It might be interesting to investigate the average kinetic energy
of an electron in a disordered metal in the presence of interaction.
At low $T$ it turns out to be
temperature independent due to the same reason as
for a quantum particle in the bath of oscillators, i.e. due to
interaction. A naive calculation along the same lines as in Section 2
yields
$\langle m{\bf \dot r^2}/2\rangle -\mu \sim 1/\tau_{\varphi}$ at $T=0$.
A more accurate analysis is beyond the frames of the present paper.
This result is just a manifestation of the well known fact:
the distribution function $n({\bf p})$ of interacting electrons
(not quasiparticles) is nonzero for
any ${\bf p}$ even at $T=0$ \cite{Lifsh}.

An obvious consequence of our results is that
the decoherence rate $1/\tau_{\varphi}$ exceeds temperature at
sufficiently low $T$. Does this fact imply the breakdown of
the Fermi liquid theory (FLT) hypothesis at such $T$? From a formal
point of view the violation of the inequality $T\tau_{\varphi} >1$
is yet insufficient for such a conclusion (see also
\cite{AW}), although it obviously does not support the FLT hypothesis either.
It was demonstrated
in Section 2 that a finite decoherence rate for a certain
variable of interest has no direct relation to the lifetime
of quasiparticles (the latter is infinite in the case of a free particle
in the Caldeira-Leggett bath). In our calculation
$\tau_{\varphi}$ determines the dephasing time for {\it real
electrons} and not for the Landau quasiparticles. Our result implies that
interacting electrons are ``bad'' particles since their
wave functions dephase even at $T=0$. The possibility to construct
``better behaving'' quasiparticles remains questionable for disordered
metals. Furthermore, even if such quasiparticles exist their properties are
completely unknown. Therefore at this stage they can hardly be used
for calculation of any physical quantity. An important advantage of
our method is that it allows for a direct calculation of measurable
quantities in terms of interacting electrons without appealing
to the Fermi liquid hypothesis.

\section{Discussion of Experiments}

Our results for the decoherence time at low $T$ turn out to be
in a remarkably good agreement with available experimental data
obtained in various physical systems.
In Ref. \cite{GZ} we have already carried out a detailed
comparison between our theory and the experimental results
\cite{Webb} obtained for 1d Au wires. The agreement within a
numerical factor of order one
was observed for all samples studied in \cite{Webb}.
Since both $\tau_{\varphi}$ and
$L_{\varphi}$ are {\it defined} with such an accuracy no better agreement
can be expected in principle.

Here we will present a comparison of our theory with two other
experiments carried out with 2d electron gas in semiconductor
structures. These systems are somewhat different from the
metallic ones mainly because of much higher effective resistance.
The parameters of the systems were chosen in a way
to realize a quasi 1d conducting system with disorder. One such
experiment was carried out by Pooke {\it et al.} \cite{Pooke}
in narrow Si pinched accumulation layer MOSFETs.
These authors studied samples with resistances ranging from
120 to 360 k$\Omega$ and observed a finite decoherence rate
$1/\tau_{\varphi}$ at all temperatures. On a log-plot
(not shown) clear signs of saturation at low $T$ are seen.
The corresponding data
for $L_{\varphi}$ \cite{Pooke} obtained for 3 samples
are presented in Fig. 2 and in the Table 1
together with our theoretical predictions. Note, that the experimental
value of $L_{\varphi}$ at $T=0$ was obtained by a linear extrapolation
$1/L^2_{\varphi}(T)=1/L^2_{\varphi}(0)+BT$
of the experimental data \cite{Pooke}.

\centerline{Table 1}
\begin{center}
\begin{tabular}{|c|c|c|c|c|}
\hline
 $V_g$, $V_p$, V
& $L_\varphi^{\rm exp}(0)$, $\mu$m & $L_\varphi^{\rm theor}(0)$, $\mu$m &
 $B^{\rm exp}$, $\mu$m$^{-2}$K$^{-1}$ &
$B^{\rm theor}$, $\mu$m$^{-2}$K$^{-1}$ \\
 \hline
 60,  0  & 1.24  & 0.78 & 5.45   & 3.8    \\
 60,  -3 & 0.85  & 0.58 & 7.83   & 5.4     \\
 75,  -6 & 0.66  & 0.53 & 6.65   & 6     \\
\hline
\end{tabular}
\end{center}

\begin{figure}[h]
\centerline{\psfig{file=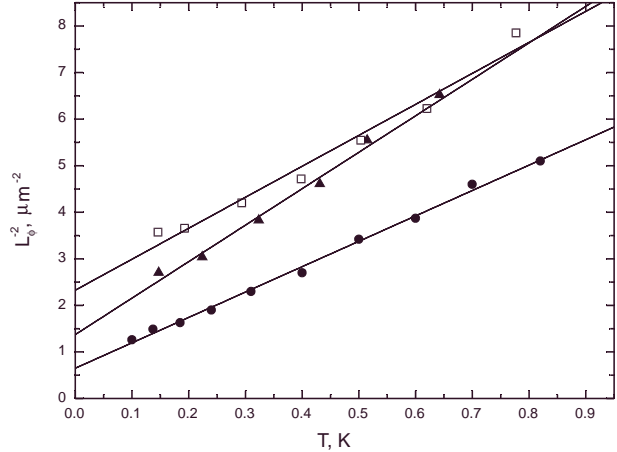,height=6cm}}
\caption{
The experimental data \protect\cite{Pooke} (symbols) fitted
to our theory (solid lines).
}
\end{figure}

The agreement between theory and experiment is within a numerical factor
of order one, i.e. again within the accuracy of the {\it definition} of
$L_{\varphi}$. Adjusting this numerical factor we observe a perfect fit
of the experimental data by our theoretical curves (see Fig. 2). As
predicted (cf. eq. (\ref{length})) at sufficiently low $T$ the value
$1/L_{\varphi}$ increases linearly with temperature.

Very recently new measurements of the dephasing time
$\tau_{\varphi}$ in quasi 1d $\delta$-doped GaAs structures
with few conducting channels were reported \cite{KGB}.
The typical effective resistance of the samples \cite{KGB}
was $6\div 30$ M$\Omega$, i.e. it was even higher than in Ref. \cite{Pooke},
the length and the width of the wires were 500 $\mu$m and 0.05
$\mu$m respectively.
Experimental results for 3 different values of the gate
voltage $V_g$ \cite{KGB} are presented in the Table 2 together
with our theoretical predictions. Again a perfect agreement
between the maximum measured value $L_{\varphi}$ and the one
derived from our theory is found for all $V_g$.

\centerline{Table 2}
\begin{center}
\begin{tabular}{|c|c|c|c|c|}
\hline
$V_g$, V &
$n$, $10^{12}$ cm$^{-2}$ &
$R$, M$\Omega$ &
$L_\varphi^{\rm exp}$, $\mu$m &
$L_\varphi^{\rm theor}(0)$, $\mu$m \\
 \hline
 +0.7  & 4    & 5.94        &  0.35  & 0.23   \\
 0     & 2.7  & $\simeq$ 18 &  0.09  & 0.08   \\
 -0.35 & 2    &$\simeq$ 36 &  0.06  & 0.04   \\
\hline
\end{tabular}
\end{center}

The temperature dependence of the data $L_{\varphi}$ \cite{KGB}
is also in excellent agreement with our predictions at all $T$,
see Fig. 3.

\begin{figure}[h]
\centerline{\psfig{file=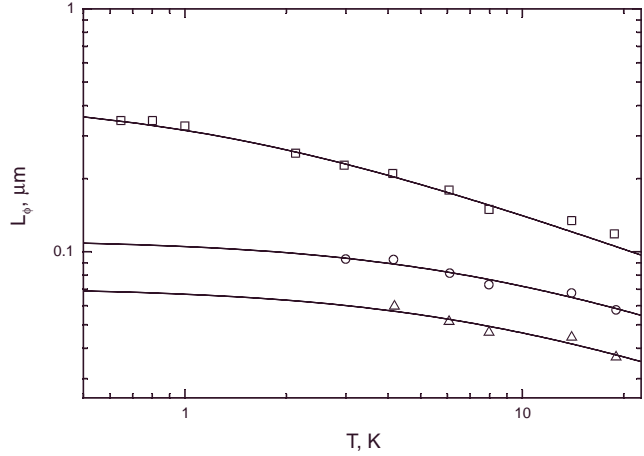,height=6cm}}
\caption{
The experimental data \protect\cite{KGB} (symbols)
fitted to our theory (solid lines).
}
\end{figure}

In Ref. \cite{KGB} the measured maximum values for
$\tau_{\varphi}$ were compared with the theoretical formula
suggested in Refs. \cite{Webb,Moh} and a discrepancy
by a factor of 50 was reported. This fact allowed the authors
to conclude \cite{KGB} that their experimental results for
$\tau_{\varphi}$ argue against the idea of decoherence by
zero-point fluctuations of the electrons. The comparison
of the data \cite{KGB} with our theoretical results clearly
demonstrates that this conclusion is simply an artefact of
the inadequate choice of a theoretical formula adopted in \cite{KGB}.
The measurements of the decoherence rate \cite{KGB}
{\it strongly support} the idea of decoherence due to intrinsic
quantum noise rather than argue against it.

In Ref. \cite{KGB} also another interesting experimental
observation was made: a crossover to a highly
resistive state was found around $T \sim 1$ K. This observation
was interpreted in \cite{KGB} as a Thouless crossover to the
regime of strong localization and was also qualified as contradicting
the very idea of quantum decoherence at $T=0$. Several comments
are in order.

(i) For the systems studied in \cite{KGB} one has
$\sqrt{N_{\rm ch}} \sim 2\div 3$. Hence, the weak localization
correction at low $T$ (\ref{delsig}) should be of
the same order as the Drude conductance (in fact, this
is exactly what was observed in \cite{KGB}) and the Thouless
crossover cannot be ruled out theoretically for such small $N_{\rm ch}$.
Also the result for $\tau_{\varphi}$ may become more
complicated for small $N_{\rm ch}$ \cite{GZ} because of a somewhat
more important
role of capacitive effects. This may in principle lead to
a relative increase of $L_{\varphi}$ (it appears, however, that
this effect is not very important in \cite{KGB}). Thus even
the presence of a Thouless crossover in the samples \cite{KGB}
would by no means contradict our theoretical results and hence
the idea of quantum decoherence at $T=0$.

(ii) The interpretation
of the observed effect as a Thouless crossover is not
quite convincing. This interpretation
is based on two reasons \cite{KGB}: (i) at the crossover $L_{\varphi}$
was found to be ``only'' $\sim 3$ times smaller than $l_{\rm loc}$ and (ii) the
resistance of a wire segment of the length $\sim l_{\rm loc}$ was found
to be $\sim 20$ k$\Omega$. The first observation does not contradict
to our theory which predicts
$l_{\rm loc}/L_{\varphi} \sim \sqrt{N_{\rm ch}} \sim 3$
at low $T$. As to the resistance of a segment $\sim l_{\rm loc}$ it
is of the order of the quantum resistance unit also at temperatures
$T \sim 3\div 10$K, i.e. well above the crossover. Hence, no definite
conclusion can be drawn.

(iii) The expression for $l_{loc}$ used in \cite{KGB} makes sense
only provided the condition $L_{\varphi}>L\gg l_{loc}$ is satisfied ($L$ is
the wire length). For $L_{\varphi} \ll L$ the wire can be viewed as
$N \sim L/L_{\varphi}$ {\it independent} samples connected in series
(in \cite{KGB} $N$ is typically of order $10^3\div 10^4$).
Since the the length of each of such samples $L_{\varphi}$ is $\sim 3$ times
{\it smaller} than the localization length $l_{loc}$ it is somewhat
naive to seriously believe that strong localization can be observed
in such samples relying only on the ``order-of-magnitude'' character
of the relation between $L_{\varphi}$ and $l_{loc}$ at the crossover.
Moreover, assuming the dependence $L_{\varphi} \propto
T^{-1/3}$ one immediately observes that the ``standard'' crossover condition
$l_{\rm loc}\sim L_{\varphi}$ would hold at temperatures $\sim 30$ times
smaller than the actual crossover temperature, i.e. at $T \sim 20 \div 30$ mK.
We see no way how the Thouless crossover can be expected above this
temperature range for the systems studied in \cite{KGB}.

(iv) The inequality $T\tau_{\varphi} \gg 1$ is violated in \cite{KGB}
at all relevant $T$: at the crossover one has $T\tau_{\varphi} \sim 0.3\div 0.5$
depending on the sample, and  $T\tau_{\varphi}$ remains of order
one even at $T \sim 10\div 20$K, i.e. well in the weak localization regime.
If one believes that the violation of the above condition signals the
breakdown of FLT due to interaction the whole
discussion of the Thouless crossover becomes pointless. An alternative
would be to acknowledge that the value $T\tau_{\varphi}$ may not
be the relevant parameter as far as FLT is concerned. But in any case
{\it real electrons} dephase and therefore can hardly be described
within the Thouless scenario of strong localization.

(v) Since $T\tau_{\varphi} < 1$ and $l_{\rm loc}$ is several times
larger than $L_{\varphi}$  at the crossover interaction definitely
plays a very important role  ``helping'' to localize electrons
instead of destroying localization. If so, why not to assume that
the whole effect is solely
due to interaction and not due to spacial disorder? For instance, it is
well known that the mobility of a quantum particle in a {\it periodic}
potential can decrease dramatically with $T$ if this particle is
coupled to a dissipative environment (see e.g. \cite{SZ}). At $T=0$
this particle can even get localized due to the effect of quantum
noise of the environment \cite{Schmid2,SZ} which completely
destroys the phase coherence of the wave function. Within this scenario
the crossover to a highly resistive state \cite{KGB} can be
considered as {\it supporting} the idea of quantum decoherence due
to intrinsic quantum noise. Thus, although a detailed interpretation
of the crossover \cite{KGB} is still an open problem, presently we see no
way to use it as an argument against quantum decoherence at $T=0$.

It was argued in \cite{KGB} that the effect of saturation
of $\tau_{\varphi}$ observed in many experiments at low $T$ can be
caused by the external microwave noise. Although filtering
of external noise is indeed a serious experimental problem
it is quite obvious that the above explanation faces
several severe problems. Without going into details let us
just indicate some of them.

Firstly, according to the arguments \cite{KGB}
the dephasing effect of the external noise may not be accompanied
by heating for low resistive samples with $R \ll 24$ k$\Omega$,
whereas in the opposite case of highly resistive samples heating
is unavoidable. It is not clear how to match this conclusion with
the experimental
results \cite{Pooke} where clear signs of saturation
of $L_{\varphi}$ at $T <$ 1K were seen for samples with resistances
up to $R \simeq 360$ k$\Omega \gg 24$ k$\Omega$.

Secondly, the formula (\ref{1})
{\it quantitatively} (within a factor of order one) describes
the low temperature value of the decoherence time $\tau_{\varphi}$
measured in different experiments in at least 10 1d
semiconductor and metallic samples. In order to interpret the results of
all these measurements in terms of external noise one should assume
that external noise always adjusts itself to a particular value
of the sample conductance (in various experiments these values
differ by many orders of magnitude) and the Fermi velocity. More than
that, the corresponding electric field produced by the
external noise inside the sample should be always of the same order
as one due to the intrinsic quantum noise. It would be interesting
to estimate the probability for such a coincidence in (at least) 10
different samples.

Thirdly, the presence of the external noise can only
be proven experimentally by making experiments with and without
necessary filtering and observing different results in these two cases.
In Ref. \cite{KGB} the external noise power was estimated, however
no evidence for its existence in experiments was presented.
In contrast, quantum noise with $\omega >T$ is well observable
reality, see e.g. \cite{QN}.

Thus the explanation of the existing experimental results in terms
of external noise turns out to be problematic. Perhaps new
experiments are needed to unambiguously rule this issue out.

\section{Conclusions}

In the present paper we have discussed the fundamental effect
of interaction induced quantum decoherence in disordered metals.

We have considered a simple model of a quantum particle interacting
with the Caldeira-Leggett bath of oscillators. An exact solution
in the case of a free damped particle demonstrates that the
off-diagonal elements of the particle density matrix decay in
the long time limit at all temperatures including $T=0$. For
a particle on a ring similar results are found.

A very transparent physical picture of the effect of quantum
decoherence due to interaction with a quantum bath of oscillators
emerges from our analysis. The interference contribution to the
return probability $W_2$ for a particle interacting with {\it one}
oscillator with a frequency $\omega$ oscillates in time and is
smaller than one for all
time moments except $t=2\pi n/\omega$ when the system returns to
its initial state. If interaction occurs with {\it infinitely
many} oscillators with a continuous distribution of frequencies
the particle will {\it never} return exactly to its initial state.
At $T=0$ the return probability will be suppressed by a factor
$$
W_2 \propto \exp (-\eta R^2),
$$
where $\eta$ is the viscosity of the environment and $R$ is the
size of the return path. This defines the typical dephasing length
$L_{\varphi} \sim 1/\sqrt{\eta }$. For a particle in a diffusive
environment the typical size of such a path grows with time
as $R \sim \sqrt{Dt}$, and the interference probability will decay
as $W_2 \propto \exp (-\eta Dt)$. This is the effect of quantum
dephasing. It has an essentially quantum mechanical nature,
therefore its existence at $T=0$ is by no means surprizing. At
nonzero $T$ the effect increases due to increasing fluctuations
in the environment.

The main features of the effect derived by means of our
simple model are reproduced within the rigorous analysis developed for
a disordered metal, in this case one should account for the
Pauli principle and the sample dimension. The electron interacts
with the fluctuating electromagnetic field produced by other
electrons. This field can be again represented as a collection of
oscillators with a somewhat more complicated spectrum than
in the Caldeira-Leggett model. Quantitatively the results will
depend on that, but the physical nature of the effect remains the same.

The effect of quantum dephasing studied here has no direct relation
to the question about the existence of ``coherent'' quasiparticles
in the problem. It is obvious that only the behavior of measurable quantities
is of physical importance. If these quantities
(e.g. the current or conductance in the case of disordered metals)
are expressed in terms of ``incoherent'' variables, quantum dephasing
yields directly measurable consequencies. A convincing illustration for
that is provided by the existing experimental data.

Our results indicate a necessity to reconsider the commonly
adopted point of view on the role of interactions in disordered
metallic systems. In particular, one arrives at the conclusion
that no electron localization takes place and low dimensional
disordered metals with generic parameters do not become insulators
even at $T=0$.

\section*{Acknowledgments}

We would like to dedicate this paper to the memory of the late Albert Schmid
whose works constitute one of the major contributions to the field. It
is a pleasure to acknowledge numerous stimulating discussions with C. Bruder,
Y. Imry, Yu. Makhlin, A. Mirlin, G. Sch\"on and P. W\"olfle in the course
of this work. We also profitted from discussions with Ya. Blanter, G. Blatter,
T. Costi, Yu. Gefen, V. Geshkenbein, V. Kravtsov, R. Laughlin, A. van Otterlo,
M. Paalanen, D. Polyakov and A. Rosch.
This work was supported by the Deutsche Forschungsgemeinschaft within
SFB 195 and by the INTAS-RFBR Grant No. 95-1305.

\end{document}